\documentclass[fleqn,twoside]{article}
\usepackage{espcrc2}

\usepackage{graphicx}
\usepackage[figuresright]{rotating}

\newcommand{\AmS}{{\protect\the\textfont2
  A\kern-.1667em\lower.5ex\hbox{M}\kern-.125emS}}

\title{Two-photon width and gluonic component of $\sigma/f_0(600)$}

\author{G. Mennessier\thanks{Email: gerard.mennessier@lpta.univ-montp2.fr}
$^{\rm{a}}$, S. Narison\thanks{Email: snarison@yahoo.fr} \address
{Laboratoire de Physique Th\'eorique et Astroparticules,
Universit\'e de Montpellier II,\\ Case 070, Place Eug\`ene Bataillon, 34095 -
Montpellier Cedex 05, France},
 and 
W. Ochs\thanks{Email: wwo@mppmu.mpg.de } \address {  
Max-Planck-Institut f\" ur Physik, Werner-Heisenberg-Institut, 
D 80805 Munich,\\ F\"ohringer Ring 6, Germany,}
}

\def\beq{%\vspace*{-0.2cm}
\begin{equation}}
\def\eeq{\end{equation}}%\vspace*{-0.2cm}}
\def\bea{%\vspace*{-0.2cm}
\begin{eqnarray}}
\def\eea{\end{eqnarray}}%\vspace*{-0.2cm}}
\def\bq{\begin{quote}}
\def\eq{\end{quote}}

\def\nnb{\nonumber}

\def\nnb{\nonumber}
\def\la{\langle}
\def\ra{\rangle}
\def\nin{\noindent}
\def\ba{\vspace*{-0.2cm}\begin{array}}
\def\ea{\end{array}}%\vspace*{-0.2cm}}

\def\b{$\bullet~$}
\def\als{\alpha_s}

\def\gg2{ \la\alpha_s G^2 \ra}
\def\gg3{g^3f_{abc}\la G^aG^bG^c \ra}
\def\ggg4{\la\als^2G^4\ra}

\def\calD{ {\cal D} }
\def\ftilde{\tilde f}

\begin{document}

\begin{abstract}
We analyse data on $\pi\pi$ and $\gamma\gamma$ scattering below 700 MeV
within an improved analytic K-matrix model. This model is based on an
effective theory with couplings between resonances, hadrons and photons.
The two-photon decay of a resonance can proceed through intermediate
transition into charged hadrons (here: $\pi^+\pi^-$) and their subsequent 
annihilation or through a ``direct'' transition into photons.
Our analysis confirms the rather large total radiative width of the 
$\sigma$ resonance which we find as $(3.9\pm0.6)$ keV but suggests
its dominance by the
$\pi\pi$ rescattering process. This process is not sensitive to the internal
structure of the resonance contrary to the direct component which we find
small,  $ (0.13\pm 0.05)$ keV, and  well consistent with the
expectations for an unmixed glueball according to the QCD sum rule
calculations. 
\vspace{1pc}
\end{abstract}

\maketitle

\section{INTRODUCTION}
%Expectations for the scalar glueball}
%%%%%%%%%%%%

The study of scalar mesons and the interpretation of the experimental
results is, despite a long lasting effort, still an active field of research
with controversies on experimental results and the theoretical
interpretation. The lightest scalar mesons have been interpreted as
conventional $q\bar q$, but also as tetra-quark or molecular states. In
addition, there is the definitive expectation within QCD of the existence of
scalar glueballs which can mix with the nearby states with quark
constituents. Depending on the mass of the glueballs, this affects strongly
the interpretation of the spectrum and leads to different scenarios.

The phenomenological analysis attempts to group the spectrum of observed 
states into appropriate $q\bar q$ or $4q$ multiplets: left over
states are possible candidates for glueballs. Also one expects 
the production of 
such states to be enhanced in ``gluon rich processes'', while they should
be suppressed in $\gamma\gamma$ reactions and corresponding rules 
hold for decays as well. 

The existence of glueballs has been predicted long ago in the early time of
QCD as consequence of the self-interaction of gluons and first scenarios
have been developed already in 1975 \cite{MIN}. Today there is agreement
on the existence of such states and the lightest state to be a scalar meson.
Quantitative results are available from Lattice QCD and from QCD sum rules
(QSSR).
The theoretical results, recalled below, suggest a light scalar glueball with
mass around 1 GeV or below. 

Among the light particles 
the $f_0(600)/\sigma$ meson could be such a gluonic resonance. 
Recent analyses of the $\gamma\gamma\to \pi\pi$ processes have extracted the
width of $f_0(600)/\sigma\to \gamma \gamma$ as: $ (4.1\pm 0.3) $
keV \cite{PENNINGTON},  $ (3.5\pm 0.7)$ or $(2.4\pm0.5) $ keV 
\cite{PENN1} to $(1.8\pm 0.4)$ keV \cite{OLLER}, while the one
from nucleon electromagnetic polarizabilities has given $(1.2\pm
0.4)~{\rm keV}$ \cite{PRADES}. 
%These widths look fairly high compared with
%most of the available theoretical estimates based on QCD dynamics. 
%For
%instance, a QSSR estimate of the $S_2\equiv (\bar uu+\bar dd)$
%$\gamma\gamma$ width gives \cite{SNA0}:
% \beq
%\Gamma (S_2\rar\gamma\gamma)\simeq 4\ga {M\over 1{\rm GeV}}\dr^3~{\rm keV}~,
%\label{widthpenn}
%\eeq
%which would correspond to a width of about 0.5 keV for $M_\sigma\approx$ 500
%MeV.\\ 
Such results have been interpreted in \cite{PENNINGTON}  as
disfavouring a gluonic nature which is expected to have a small coupling to
$\gamma\gamma$ \cite{VENEZIA,BN,SNG,ACHASOV,BARNES,SNA0}.

In our recent paper \cite{MNO} a resolution of this conflict has been
suggested. From the analysis  of the
$\gamma\gamma\to \pi\pi$
processes in the low energy region below 700 MeV we conclude that they are
dominated by the coupling of the photons to charged pions and their
rescattering, which  therefore can hide any direct coupling of the photons to
the scalar mesons. 
Some first results from this study have been presented elsewhere \cite{PETER}.
%%%%%%%%%%%%%%%%%%%%%%%%%%%%%%%%%%%%%%%%%
\section{A LIGHT $0^{++}$ GLUEBALL}
%\vspace*{-0.3cm}
There are several results suggesting a light scalar glueball:

{\it \b Lattice QCD}. 
%%%%%%%%%%%%%%%%%%%%%%%%%%%%
% \nin
 Calculations in the  simplified 
world without quark pair creation (quenched approximation) find the
lightest state at a mass around 1600 MeV 
(recent review in Ref. \cite{latticerev}). These findings
 lead to the construction of models
where the lightest glueball/gluonium mixes with other mesons in the
 nearby mass range
of around 1300-1800 MeV. % (see, for example, \cite{CLOSE}). 
However, recent results beyond this quenched approximation    
\cite{MICHAEL} indicate the lightest state with a large gluonic
 component
to fall into the region around 1 GeV, and therefore, 
a scheme based on the mixing of meson states
with all masses higher than 1300 MeV could be insufficient to
represent the gluonic degrees of freedom in the meson spectrum.
Further studies concerning the dependence on lattice spacings and the quark
mass appear important.

%%%%%%%%%%%%%%%%%%%%%%%%
{\it \b QCD spectral sum rules (QSSR) and Low-energy theorems (LET)}.
 %\label{qss}
%%%%%%%%%%%%%%%%%%%%%%%%
%%%%%%%%%%%%%%%%%%%%%%%%
%{\it QSSR and LET} - %\label{qss}
%%%%%%%%%%%%%%%%%%%%%%%%
These approaches have given quantitative estimates for the
mass and decay properties of the gluonic bound states. In particular, 
in a combined analysis of
subtracted and unsubtracted sum rules a low mass for the bare
unmixed gluonium state is obtained \cite{VENEZIA,SNG}:
\beq
M_{\sigma_B}\simeq (0.95\sim 1.10) ~{\rm GeV}~,
\label{eq:msigmab}
\eeq
besides the heavier state around $M_G\simeq (1.5\sim 1.6)~{\rm GeV}$. 
These masses are similar to the unquenched and quenched lattice results
respectively, quoted above. 
The hadronic width in this approach is obtained rather large:
\beq
\Gamma_{\sigma_B\to\pi^+\pi^-}\simeq0.7~ {\rm GeV}~,
  \label{eq:scalarwidth}
\eeq
whereas the radiative decay width is found small: 
\beq
\Gamma_{\sigma_B\to\gamma\gamma}
\simeq (0.2\sim 0.6)~{\rm keV}~.
\label{eq:radwidth}
\eeq
% \cite{SNG0,SNG1,SNG} and of some essential
% features of its branching ratios \cite{VENEZIA,SNG,SN06,NSVZ,LANIK}. 
Indeed, this width is considerably smaller than the value of about 5 keV,
which is expected for a state
 $S_2\equiv (\bar uu+\bar dd)$ with mass of 1 GeV \cite{SNA0}. 
On the other hand, QSSR predicts, for a
four-quark state having the same mass of 1 GeV, a
$\gamma\gamma$ width of about 0.4 eV \cite{SNA0}. 

%%%%%%%%%%%%%%%%
{\it \b Phenomenological studies}.  
%%%%%%%%%%%%%%%%
A full understanding of the scalar meson spectrum is required
if the glueball is to be found as a left over state after the identification
of flavour multiplets.
Several schemes exist, motivated by the quenched lattice result, 
 where the extra gluonic state is assumed to mix into 
the three isoscalars $f_0(1370),\ f_0(1500)$ and $f_0(1710)$ \cite{CLOSE}. 
In an
alternative approach \cite{OCHS},  the 
lightest $q\bar q$ multiplet is formed
from $f_0(980), f_0(1500), K^*_0(1430)$ and $a_0(1450)$, and the glueball
is identified as 
the broad object at smaller mass represented by $f_0(600)$; 
it dominates $\pi\pi$ scattering near 1 GeV but extends from threshold up
to $\sim 1800$ MeV (see also ref. \cite{ochs06}).
The appearence of this broad object 
in most gluon rich processes was considered 
in support of this hypothesis. In this analysis of the spectrum, 
results from elastic and
inelastic $\pi\pi$ scattering as well as from $D$, $B$ 
and $J/\psi$ decays have been considered \cite{OCHS,mo2}.

%%%%%%%%%%%%%%%%%%%%%%%%%%%%%%%%%%%%%%%%%%%%%%%%%%%%%
%\section{Model for $\gamma\gamma$ from $\pi\pi$ processes}%\vspace*{-0.25cm
%%%%%%%%%%%%%%%%%%%%%%%%%%%%%%%%%%%%%%%%%%%%%%%%%%%%%

%%%%%%%%%%%%%%%%%%%%%%%%%%%%%%%%%%%%%%%%%%%%%
%\vspace*{-0.3cm}
\section{THE DIRECT CONTRIBUTION}
%%%%%%%%%%%%%%%%%%%%%%%%%%%%%%%%%%%%%%%%%%%%%
The amplitudes for the reactions  
$\gamma\gamma\to \pi\pi$ are largely determined by
 elastic $\pi\pi$ scattering and likewise in
the case of multi-channel processes.
First, there are the extended unitarity relations
corresponding to the Watson theorem for the single channel; secondly, there
are dispersion relations where the necessary subtraction terms
generate polynomial ambiguities
\cite{OMNES}. %for the single- and multi-channel cases.
% respectively.
%%%%%%%%%%%%%%%%%%%%%%%%%%%%%%%%%%%%
%\vspace*{-0.3cm}
%\subsection*{\b The analytic K-matrix model}
%%%%%%%%%%%%%%%%%%%%%%%%%%%%%%%%%%%%
% \nin
This general formalism has been 
applied by Mennessier \cite{mennessier},  
to obtain the electromagnetic
processes $\gamma\gamma\to \pi\pi,K\bar K$, given the strong processes
$\pi\pi\to \pi\pi,K\bar K$. The latter ones are represented by a K
matrix model which represents the amplitudes by a set of resonance poles.
In that case, the dispersion relations in the multi-channel case 
can be solved explicitly, which is not possible otherwise. 
This  model can be reproduced by a set of Feynman diagrams, including 
resonance (bare) couplings to  $\pi\pi$ and $K\bar K$, 
 and 4-point $\pi\pi$ and $K\bar K$ interaction vertices.  
A subclass of bubble pion
 loop diagrams including resonance poles in the s-channel are resummed
 (unitarized Born).
%The model also includes contributions from the exchange of vector mesons
%in the t-channel which become important at the higher energies above 700
%MeV.
%The $\gamma\gamma$ scattering amplitudes also fulfill the constraints at
%$s=0$ required by the Born approximation. 

The radiative width of the resonances cannot be predicted due to the
polynomial ambiguity, which can be taken into account by introducing as free
parameters the ``direct couplings'' of the resonances to $\gamma\gamma$ in
an effective interaction vertex. In the present analysis we have extended
the model by the introduction of a {\it shape function} which takes
explicitly into account left-handed cut singularities for the strong
interaction amplitude. This allows a more flexible parametrisation of the
$\pi \pi$
% scattering
data at low energies and 
%can be used later to 
improves the high energy behaviour.
%as well. 
%Next we discuss the features of the model at low and at high energies.
% within the context of this model.
The model shows different characteristics at low and high energies:
%%%%%%%%%%%%%%%%%%%%%%%%%
%\vspace*{-0.3cm}

{\it \b Low energy limit: rescattering}. 
%-------- subsection should be reduced further ------------
% \nin
%%%%%%%%%%%%%%%%%%%%%%%%%
A striking feature of the low energy $\gamma\gamma\to \pi\pi$ scattering is the
dominance of the charged over the neutral $\pi\pi$ cross section by an order
of magnitude which can be explained by
the contribution of the pion-exchange Born term in 
$\gamma\gamma\to\pi^+\pi^-$. In the process $\gamma\gamma\to
 \pi^0\pi^0$,
the photons cannot couple ``directly'' to $\pi^0\pi^0$ but through
 intermediate charged
pions and subsequent rescattering with charge exchange. 
This feature is realized in the analytic model considered here 
\cite{mennessier} which has it in common with the Chiral perturbation
theory \cite{donoghue}.
%%%%%%%%%%%%%%%%%%%%%%%%%%
%\vspace*{-0.3cm}

{\it \b High energy limit: direct production}.
%%%%%%%%%%%%%%%%%%%%%%%%%
% \nin
%Wheareas at low energy the photon interacts by its coupling to the charged
%pions, the situation becomes different 
On the other hand, at high energies the photon can
resolve the constituents of the hadrons. An example is the production of
$f_2(1270)$ in $\gamma\gamma\to \pi\pi$. The analytic model 
\cite{mennessier} predicts a
decreasing cross section for this process with pion exchange 
reaching $<10\%$ of the observed cross section at the peak of the $f_2$.
%After inclusion of vector exchange, this contribution doubles. 
%Because of
%form factor effects involved in the photon coupling to hadrons, these
%predictions from point like hadrons are to be taken rather as upper limits. 
These considerations require the need of 
a ``direct coupling'' of $f_2\to \gamma\gamma$.
Indeed, it is well known that the radiative decays of the tensor mesons
$f_2,f_2',a_2$ are well described by a model with direct coupling to the
$q\bar q$ constituents according to the $SU(3)$ structure of the nonet 
with nearly ideal mixing (see e.g. Ref. \cite{ARGUS}).  
%Another example of a direct process is the production of $\rho$ mesons
%n the process $\gamma p\to \pi^+\pi^-p$ by direct VMD coupling, 
%hich is found much larger
%han the rescattering contribution from the ``Drell-S\"oding'' background
%process \cite{SLACRHO}.  
We therefore interpret the direct terms in the $\gamma\gamma$ processes as
originating from the coupling  of the photon to the partonic constituents
of the resonances (such as $q\bar q$, $4q$, $gg$,$\ldots$). 
%%%%%%%%%%%%%%%%%%%%%%%%%
%\vspace*{-0.3cm}
\section{ANALYTIC K-MATRIX MODEL}
%Direct  and rescattering couplings  of $\sigma$}\vspace*{-0.25cm}
%%%%%%%%%%%%%%%%%%%%%%%%%
% \nin
In the application of this model \cite{mennessier}, 
we first obtain  a suitable
K-matrix 
parametrisation of the $\pi\pi$ scattering data, and then determine the
direct coupling by comparing the model with the $\gamma\gamma$ results.
In the present analysis,
%discussion of $\pi\pi$ scattering, 
we restrict ourselves to the low mass region
below 700 MeV where we neglect vector and 
axial-vector exchanges \cite{mennessier}, 
inelastic channels and D-waves. Furthermore, we assume a pointlike pion-photon
coupling, which is expected to be a good approximation in the region where we are working.
%%%%%%%%%%%%%%%%%%%%%%%%%%%%
%\vspace*{-0.3cm}
%\subsection*{\b Amplitude for elastic $\pi\pi$ scattering}
%%%%%%%%%%%%%%%%%%%%%%%%%%%%
% \nin

We apply the analytic model \cite{mennessier}, but we introduce 
a {\it shape function} $f_0(s)$ which  multiplies the $\sigma\pi\pi$ coupling. 
The real analytic function $f_0(s)$ is regular for  $s > 0$
and has a left cut for  $s \le 0$. For our low energy approach, a convenient
approximation, which allows for a zero at $s=s_A$
and a pole at $\sigma_D>0$ simulating the left hand cut, is: 
\beq
f_0(s)=\frac{s-s_{A0}}{s+\sigma_{D0}} \label{formfactor}~.
\eeq 
For simplicity, we don't include the 4-point coupling
term.
The unitary 
$\pi\pi$ amplitude $T_\ell^{(I)}$ for the isospin $I=0$ S-wave
is written as:
\beq
  T_0^{(0)}(s) = \frac{G f_0(s)}{s_R-s - G \ftilde (s)} = \frac{G f_0(s)}{\calD(s)}~, 
\label{tpipi}
\eeq 
where  $G=g_{\pi,B}^2$ is the bare coupling squared. Unitarity determines
the imaginary part, while  a dispersion relation subtracted at $s=0$ 
determines the form of $\tilde f(s)$ \cite{MNO}. 
% $ \rho(s) =({1 - 4 m^2_\pi/s})^{1/2}$;  and:
%The amplitude near the pole $s_0$ where $ {\cal D}(s_0)=0$ and
%$\calD(s)\approx \calD'(s_0) (s-s_0)$ is:
%\beq
%  T_0^{(0)}(s)\sim \frac{g_\pi^2}{s_0-s}; \qquad g_\pi^2=\frac{G
%f_0(s_0)}{-\calD'(s_0)}~.
%\label{eq:gpi2}  
%\eeq

%%%%%%%%%%%%%%%%%%%%%%%%%%%%
%\vspace*{-0.3cm}
%\subsection*{\b Amplitudes for $\gamma\gamma\to \pi\pi$ scattering}
%%%%%%%%%%%%%%%%%%%%%%%%%%%%
% \nin
Using the
 S wave amplitude in Eq.~(\ref{tpipi}) we derive the amplitude $T_\gamma^{(I)}$ for 
the electromagnetic process for isospin $I=0$ as:
\beq
 T_{\rm \gamma}^{(0)} = 
  \sqrt{\frac{2}{3}} \alpha \left(f_0^B + G\ \frac{\ftilde_0^B}{\calD}\right) + 
  \alpha \frac{\cal P}{\calD}~. \label{elmamp} 
\eeq
The first contribution comes from the Born term for
$\gamma\gamma\to \pi^+\pi^-$, the second one from the $\pi\pi$ rescattering and
the third one represents the polynomial contribution 
${\cal P} = s F_\gamma\sqrt{2}$ which reflects the ambiguity from the
dispersion relations and  represents the direct
coupling of the resonance to $\gamma\gamma$. The residues 
 at the pole $s_0$ of the
rescattering  
and direct contributions to $ T^{(0)}_\gamma$ in Eq.~(\ref{elmamp})
 determine the respective branching ratios.  
Similarly, we construct
the  $I=2$ $S$-wave amplitude $T^{(2)}_0$ 
and  $T^{(2)}_\gamma$ as well as the 
cross sections for the $\pi\pi$ and $\gamma\gamma$ 
scattering processes \cite{mennessier}.

 %%%%%%%%%%%%
% \vspace*{-.5cm}
\begin{figure}[t]
\begin{center}
\includegraphics[width=7.2cm]{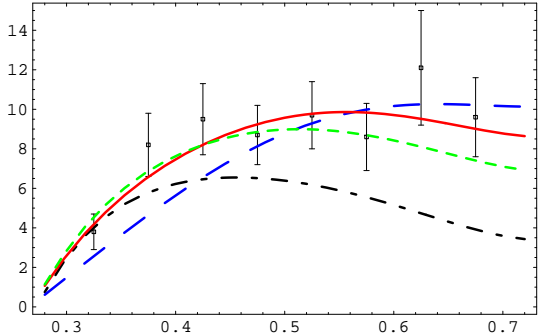} %original width 8cm
\vspace*{-0.5cm}
\caption{%\footnotesize  a) 
 $\pi^0\pi^0$ cross-section [nb]  versus $\sqrt{s}$ [GeV]
and fit by analytical model \protect\cite{mennessier}: $F_\gamma=0$ 
(dot\-dashed);
 $F_\gamma=-0.09$: I=0 (large dashed), I=0 +2 (continuous); 
$F_\gamma=-0.07$: I=0+2 (small da\-shed). 
Data from Crystal Ball \cite{cball} ($|\cos\theta| \leq 0.8$). 
 %b) Corresponding total $\chi^2$ for 8 number of data points versus $F_\gamma$.
 }
 \label{fig:fitneutral}
\end{center}
\vspace*{-1.cm}
\end{figure}
%%%%%%%%%%%
 %%%%%%%%%%%%
\begin{figure}[tb]
\begin{center}
\includegraphics[width=7.2cm]{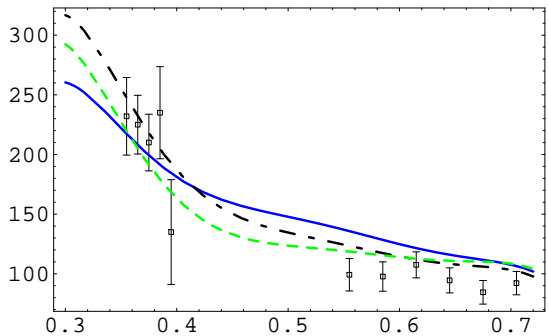}
\vspace*{-0.5cm}
\caption{%\footnotesize 
Same as in Fig. \ref{fig:fitneutral} but for $\pi^+\pi^-$: 
$F_\gamma=0$ (dot-dashed);
 $F_\gamma=-0.07$ and I=0+2 (small dashed). The continuous line corresponds
to the non-unitarized Born amplitude with $F_\gamma=0.$ 
Data are MARKII \cite{MARK2} ($|\cos\theta| \leq 0.6$).
% b) Total $\chi^2$ from a simultaneous fit of the $\pi^+\pi^-$ and $\pi^0\pi^0$ total cross-sections shown in the previous figures for 19 number of data points
%versus $F_\gamma$. 
}
\label{fig:fitcharged}
\end{center}
\vspace*{-1.cm}
\end{figure}
%%%%%%%%%%%
 %%%%%%%%%%%%
%\begin{figure}[hbt]
%\begin{center}
%\includegraphics[width=4.5cm]{fig00.eps}
%\includegraphics[width=6cm]{chi2.eps}
%\includegraphics[width=6cm]{fig++.eps}
%\vspace*{-0.4cm}
%\caption{\footnotesize Total $\chi^2$ from a simultaneous fit of the $\pi^+\pi^-$ and $2\pi^0$ total cross-sections shown in the previous figures. }
%\label{fig:chi2tot}
%\end{center}
%\vspace*{-0.8cm}
%\end{figure}

%%%%%%%%%%%%%%%%%%%%%%%%%%%%%%%%
%\vspace*{-0.3cm}
\section{RESULTS OF THE ANALYSIS}
%%%%%%%%%%%%%%%%%%%%%%%%%%%%%%%%
 \nin
We first determine the parameters of the model for $\pi\pi$ scattering below
700 MeV from the best approximation of our formula to the phase shifts
$\delta_0^{(0)}$ and $\delta^{(2)}_0$ obtained in the Roy equation analysis 
in \cite{ccgl}.
%which takes into account the high-energy behaviour suggested in \cite{YND}. 
For both channels $I=0$ and $I=2$ we fit 4 parameters each. We find the pole mass:
\beq
 M_\sigma\simeq (422-{\rm i}~290)~{\rm MeV} ~%~~g_\pi \simeq  0.06 -{\rm i}~0.50~{\rm GeV},
%\label{eq:pipiwidthfit}
\label{eq:gpi}
\eeq
which is close to the masses in other recent determinations
 (441 - i 272) MeV  \cite{leutwyler}\  or
(489 - i 264) MeV  \cite{YND}.

Given the $\pi\pi$ amplitude,  we can predict the cross sections for
$\gamma\gamma \to \pi\pi$ where the only free parameter $F_\gamma$ is
related to the strength of the direct coupling $\sigma \to \gamma \gamma$.
The fit of the model 
to the data from Crystal Ball ($\pi^0\pi^0$) \cite{cball}
and MARK-II ($\pi^+\pi^-$) \cite{MARK2} collaborations is shown in
Figs.~\ref{fig:fitneutral} and \ref{fig:fitcharged} for the cases with 
direct contribution and
without (``unitarised Born'' : $F_\gamma=0$).
%%%%%%%%%%%
We obtained excellent fits to the neutral pion data with $F_\gamma=-0.090$.
A fit to both channels yields  $F_\gamma=-0.070$. Then the fit
in the charged channel 
deviates from the data at high-mass; the large systematic errors 
and the absence of data points in the region 0.40 to 0.55 GeV 
 do not permit a good understanding of this channel.  
We take as a final estimate 
$F_\gamma \simeq - (0.080\pm 0.014)$.
%where we consider that the uncertainty in the fitting procedure 0.010 is expressed by this range.  We have
%added linearly  the systematics of the method to be about 5\% , which we have estimated from the deviation of the mass and width determinations in Eq. (\ref{eq:pipiwidthfit}) from the ones \cite{leutwyler} in Eq. (\ref{sigma}), as we have used their parametrization for fixing the parameters of the model.
From the residues at the pole $s_0$ and
$\Gamma_{\sigma\to \pi\pi}= 2 {\rm Im} M_\sigma$,  
%\vspace*{0.4cm}
%\\
we can deduce the ``partial" $\gamma\gamma$ widths at the complex pole :
\bea
\Gamma_{\sigma\to\gamma\gamma}^{\rm dir}&\simeq& (0.13\pm 0.05)~{\rm keV}~,\nnb\\
\Gamma_{\sigma\to\gamma\gamma}^{\rm resc}&\simeq& (2.7\pm 0.4)~{\rm keV}~,
\label{eq:gamsigma}
\eea
%\vspace*{-0.2cm}
 and the total $\gamma\gamma$ width (direct + rescattering):
% \vspace*{0.3cm}
 \beq
 \Gamma_{\sigma\to\gamma\gamma}^{\rm tot}\simeq (3.9\pm 0.6)~{\rm keV}~.
 \label{radtot}
\eeq
%There are also theoretical uncertainties such as the presence of other
%exchange processes which are not considered here.
%and the
%inclusion of the higher mass resonances which may affect these values.

%%%%%%%%%%%%%%%%%%%%%%%
%\vspace*{-0.3cm}

Our result for the direct coupling is
comparable with our previous value \cite{PETER} obtained 
%from the $\gamma\gamma\to\pi^0\pi^0$ data alone and 
without using the shape function
$f_0(s)$, but not as small as in a parallel analysis \cite{ACHASOV07}. 
Our  total $\gamma\gamma$ 
width is compatible with the range of values
%$(1.2 \sim 3.2)$ keV obtained in \cite{PETER} using a Breit-Wigner parametrization of the $\gamma\gamma\to \pi^0\pi^0$ data.
obtained previously
\cite{PENNINGTON,PENN1,OLLER,PRADES}.
These analyses, however, don't separate the direct term.

The results from QSSR/LET 
are obtained in the physical region
and can better be compared with experiment
%more appropriate
by using the corresponding results for
 the ``visible meson'' on the real axis instead of the results at the
complex pole:
this can be,  either the Breit-Wigner mass and width (see \cite{PETER}) or   
the ``on shell'' mass and width (see e.g. \cite{SIRLIN}),  which are obtained   
for the mass $M^{\rm os}_\sigma$ where the real part of the propagator vanishes 
${\rm Re} {\cal D}({(M^{\rm os}_\sigma})^2)=0$. For this definition, we
obtain:
\bea
M_\sigma^{\rm os}\approx 0.92~{\rm GeV},&& \Gamma_{\sigma\to\pi\pi}^{\rm os}\approx
1.02~{\rm GeV},\nonumber\\ 
\Gamma_{\sigma\to \gamma\gamma}^{\rm os,dir}
&\approx& (1.0\pm 0.4)~{\rm keV}.
\label{eq:onshell}
\eea
As the model is extrapolated here towards
energies beyond its validity ($M\sim 1$ GeV), we 
consider these results as a crude approximation.
%and will only serve as a guideline.

%%%%%%%%%%%%%%%%%%%%%%%%%%%%%%%%%%%%%%
%\vspace*{-0.3cm}
\section{COMPARISON WITH QSSR/LET}% RESULTS} \label{qssr2}%\vspace*{-0.25cm}
%%%%%%%%%%%%%%%%%%%%%%%%%%%%%%%%%%%%%%
% \nin
 
 One can notice that the QSSR and LET predictions for the mass and hadronic 
widths of a low mass gluonium $\sigma_B$ in Eqs. (\ref{eq:msigmab}) and 
(\ref{eq:scalarwidth})
are in remarkable agreement with the results in Eq. (\ref{eq:onshell}) 
for an on-shell resonance.
The direct $\gamma\gamma$ coupling, which can reveal the   
photon coupling to the intrinsic quark ($q\bar q,\ 4q,\ldots$) or gluon
($\ gg,\ldots$) constituent structure of the resonance, can be 
related to the QSSR/LET evaluations of its width through quark or
gluon loops as in case of the quark triangle for the pion or the
$f_2(1270)$. Our $\sigma\to\gamma\gamma$ width in 
Eq. (\ref{eq:onshell}) (also the  
pole value in Eq. (\ref{eq:gamsigma})) are consistent with the small value
predicted in Eq. (\ref{eq:radwidth}).
This ``overall agreement'' favours a large gluon component in the
$\sigma/f_0(600)$ wave function. It
disfavours, in particular, a $\bar qq$  interpretation because of 
the larger radiative width to be expected ($\sim 5$ keV). On the other hand,
a four-quark interpretation would require a smaller radiative width and is
disfavoured as well by our result in Eq. (\ref{eq:gamsigma}) 
(at the level of 2.5 $\sigma$) . 

%---------------- studied up to here ---------------
%%%%%%%%%%%%%%%%%%%%%%%%%%%
%\vspace*{-.35cm}

Improvements of our estimates need more precise data below 700 MeV, 
which could be provided by the KLOE-2 experiment in the future \cite{BLOISE}.
Furthermore, an extension of the analysis to higher energies will be
important. 

% \vspace*{-0.7cm}
\section*{ACKNOWLEDGEMENT}
%\vspace*{-0.25cm}
%\nin
%%%%%%%%%%%%%%%%%%%%
We thank Peter Minkowski for discussions and 
the collaboration in an early phase of this work.
%\vfill \eject
%%%%%%%%%%%%%%%

\end{document}